# An Empirical Look at Neutrino Oscillations[*]

*by John Michael Williams*

**jwill@AstraGate.net**
*Markanix Co.*
P. O. Box 2697
Redwood City, CA 94064

[*] online at *arXiv*, physics/0107017

## Abstract

The data supporting neutrino oscillations are reexamined empirically, ignoring the phase space of the usual theory. An absolutely minimum description can be constructed easily without assuming oscillations. An empirical fit to a simplified but representative set of neutrino problem data may require only two free parameters; the usual oscillation theory requires as many as four.

Free parameters used were total mass in range of a propagating neutrino, distance travelled, and a detector profile parameter.

The usual oscillation theory appears overcomplicated and inefficient. Even if future data do not demand it be complicated further, it should be abandoned for something better.

# Introduction

Because the usual neutrino oscillation theory appears to be flawed [1], the question arises as to what actually are the data that a theory of neutrino oscillations should have to explain?

As of today, the problem(s) to be explained fall into three coarse categories: Solar observational data, atmospheric observational data, and experimental data. We propose to look at the following:

- For the ***solar*** data, from Super-KamiokaNDE (Super-K; [2]) and Sudbury Neutrino Observatory (SNO; [3]);

- for the ***atmospheric*** data, from Super-K [4]; and,

- for the ***experimental*** data, from Los Alamos Liquid Scintillator Neutrino Detector (LSND; [5]) and KEK-to-Kamioka (K2K; [6]).

Other, earlier studies (see [7; 8]) will be ignored here because the ones chosen seem to represent them reasonably well. If we wish to understand neutrinos, we must be able to describe simply at least the broadest aspects of the data in the categories above.

Naively, decay or scattering might be advanced as explanations of the problems above, because these processes explain other particle phenomena well. However, in the past few years, as data became available, no theory based on neutrino decay or scattering could be accepted comfortably, so a theory of flavor oscillation was advanced [7 - 9]. Unfortunately, flavor oscillations, even if derived theoretically in a meaningful way, seem to require different parameters for each of the three problem categories above; so, it remains very unclear whether there has been any progress at all in explaining the data.

  We thus propose to ignore the usual neutrino oscillation theory, along with its phase space for representing the data: We seek to learn how many degrees of freedom would remain, and where.

## Data Summary

  Although tau neutrinos ($n_t$) have been identified in an emulsion detector [10], and may have been recorded in a water Cerenkov detector [11], neither they nor the hypothetical sterile neutrinos ($n_s$) have been shown to contribute measurably to any of the effects listed above, so we shall not be concerned with these specific flavors here. We thus consider only electron neutrinos ($n_e$) and muon neutrinos ($n_m$). Also, without prejudice on the issue of whether neutrinos might be Dirac or Majorana particles, we hereafter ignore the difference between neutrinos and antineutrinos, considering flavor as the sole identifying feature.

  Here are the data to be explained:

**Table 1. Coarse description of the neutrino data in the literature. See text for references.**

| Neutrino Flavor | Effect (*Flux*) | Problem Category |
|---|---|---|
| electron | ~45% disappearance | Solar (Super-K; SNO) |
|  | ~4.5% appearance | Experimental (LSND) |
| muon | ~35% disappearance | Atmospheric (Super-K) |
|  | ~25% disappearance | Experimental (K2K) |
|  | ~0.25% disappearance | Experimental (LSND) |

### *Some Apology for Our Choice of Explanation*

  We have tried to avoid previous terminology by describing the effect of flavor change as a flux deficit (or surfeit) in the initial flavor.

  The "effects" in Table 1 most certainly have been found to depend on neutrino energy; however, in terms of the data, which is to say, the results reported at any of the neutrino detectors in Table 1, this is not a free parameter but a measurement. We prefer to avoid the important energy variable to avoid redundancy with the usual analysis, and so we have globbed it into Table 1 indiscriminately. On the other hand, distance (the usual $L$) can be varied by source-detector manipulations and clearly differs among the effects listed above; so, we have used it as a parameter in our attempts at empirical description.

  For present purposes, we shall not be concerned with errors or differences of interpretation of the effects in Table 1 within ten percent or so.

  LSND is listed twice, because observation of an appearance is actual evidence of the complementary disappearance. The 4.5% appearance for LSND was estimated from about 88 excess $n_e$ [5] in about 2000 total $n_e$. Although not reported as a disappearance, the [5] calculated 0.26% oscillation rate of $n_m$ to $n_e$ is as strong evidence of disappearance as any of the other in the neutrino oscillation literature.

Strangely, although oscillations can be proven to differ from decay or absorption only by appearance of a new flavor component, the only study reporting an appearance, LSND, often is ignored by oscillation theorists. There are systematic reasons to doubt the LSND numbers, but they aren't so good as the theoretical reasons [1] for rejecting oscillation theory as such. So, we try to consider all possibilities below.

## A Minimal Nonoscillation Explanation

In Table 1, we find that the data describe (*a*) a disappearance approaching half of all neutrinos expected from the core of the Sun, (*b*) a smaller disappearance for atmospheric neutrinos passing through a diameter of the Earth, and (*c*) a still smaller disappearance of neutrinos passing through about 250 km of the Earth. We also find (*d*) a tiny disappearance, as well as appearance, of neutrinos after passage through about 30 m of dense matter.

There is no reasonable *prima facie* justification in these data to assume oscillations; but, the degree of flavor discrepancy seems related to the amount of matter traversed. So, let us start by describing the effects by amount of matter.

Discounting gravity (which should have negligible influence on the effects tabulated), neutrinos are not known to interact with matter except by the weak force, which has a range limited approximately by $1/m_W^2$ and surely less than $10^{-17}$ m. The mass in the volume of a circular cylinder with radius $10^{-17}$ m around the expected neutrino path, for each effect in Table 1, is calculated in Table 2:

Table 2. Amount of matter (mass) in weak range of a propagating neutrino, for the effects identified in Table 1.

| Effect | Path Length (m) | Average Matter Density ($kg/m^3$) | Mass in Range (kg) | Mass in Range ($eV/c^2$) |
|---|---|---|---|---|
| solar $n_e$ | $7 \cdot 10^8$ | $1.4 \cdot 10^3$ | $3 \cdot 10^{-22}$ | $2 \cdot 10^{-3}$ |
| atmospheric $n_m$ | $1.3 \cdot 10^7$ | $5.5 \cdot 10^3$ | $2.2 \cdot 10^{-23}$ | $1.4 \cdot 10^{-4}$ |
| K2K $n_m$ | $2.5 \cdot 10^5$ | $2.5 \cdot 10^3$ | $2 \cdot 10^{-25}$ | $1.2 \cdot 10^{-6}$ |
| LSND $n_m$ or $n_e$ | $3.0 \cdot 10^1$ | $6 \cdot 10^3$ | $4.7 \cdot 10^{-29}$ | $3 \cdot 10^{-10}$ |

Because the Super-K day/night effect for solar neutrinos has been shown negligible, we have ignored matter in the Earth in the first row in Table 2.

Looking at all the data in Table 2, a logarithmic function makes for a reasonable fit, if we consider as a single datum the expectancy of the two LSND results. A plot is shown in Fig. 1:

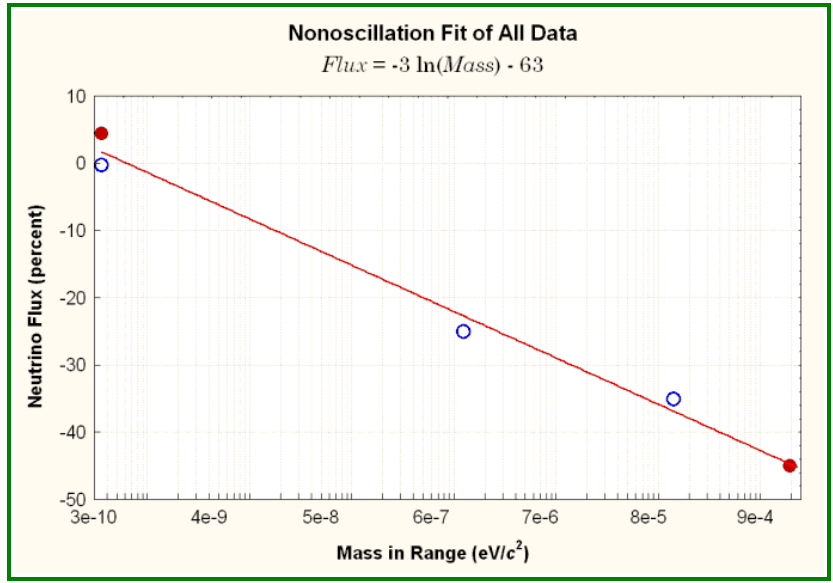

**Figure 1.** Fit of $Flux = A \cdot \ln(Mass) + B$ to the data in Table 2. Negative flux means loss of flavor (disappearance). The solid points represent $n_e$ data.

Solving
$$Flux = -3 \cdot \ln(Mass) - 63 \tag{1}$$

for mass in range, we get,
$$Mass = 7.5 \cdot 10^{-10} \exp(-Flux/3), \tag{2}$$

which is replotted in Fig. 2:

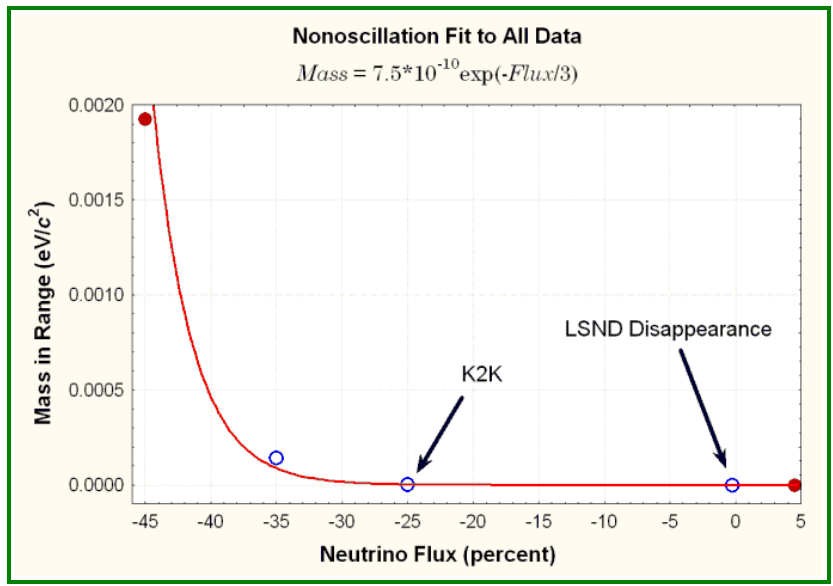

**Figure 2.** Exponential curve fit of $Mass = A \cdot \exp(B \cdot Flux)$ to the data in Table 2. Solid points represent $n_e$ data.

　　The exponential representation is somewhat insensitive to the current dataset in the variable of interest, *Flux* in the initial flavor. If future data involving large disappearance values became available, this representation would increase greatly in value.

Notice that (1) may be interpreted as an integral equation and rewritten,

$$d(Flux) = -3 \cdot \frac{d(Mass)}{Mass}; \text{ or,} \tag{3}$$

$$\frac{d(Flux)}{d(Mass)} = -3 \cdot Mass^{-1}, \tag{4}$$

the negative sign representing loss of otherwise accountable neutrino current or flux from the source.

Physically, what might Eq. (1) or (4), or Fig. 1, mean? (Empirically, it need not mean anything, so long as it works; that is what *empirical* means).

First, Eq. (1) requires some kind of renormalization, because as the amount of matter traversed approaches 0, an infinity of neutrinos can not be allowed to appear. We shall ignore this as digressive in the present context, and cut off the function where the data end.

Disregarding singularities, a somewhat *ad hoc* interpretation of Eq. (1) or Fig. 1 might be that, for substantial amounts of matter, neutrinos will disappear more, the greater the amount of matter they traverse. Of itself, this does not suggest an oscillation or resonance process; rather, decay [15, 16, 19], coherence damping [17], scattering, or absorption would seem most consistent with the empirical data. A linear, first-order process might be postulated.

The difficulty with this simple assignment of physics to the empirical representation arises in Fig. 1 where the amount of matter traversed is very small, approaching 0. Here, neutrinos of a given flavor will appear in greater numbers than they were believed to have been created.

Possibly, to avoid flavor-changing oscillations (as is the present purpose), one might postulate stimulated emission by neutrinos of the same flavor which had not yet traversed much matter. So, this effect might die off after a short distance in matter, because of early loss of coherence? Stimulated emission has been suggested before [14] and is no more radical an idea than the Mikheyev-Smirnov-Wolfenstein (MSW) hypothesis [9, 8.4], which also depends on a generic (universal) matter effect, there set proportional to the weak coupling constant.

Another idea might be regeneration of neutrinos by unexpectedly high early forward scattering cross-section, or almost any other interaction [18] changing a neutrino to a different flavor. In either case, though, some change of law has to occur after propagation over a short distance, so that positive *Flux* does not occur any more after traversal of a certain amount of matter.

## A Minimal Oscillation Explanation

There was no obvious need for oscillatory behavior in the direct fits above. Let us now look at the problem on the assumption of *Flux* oscillations.

We start by revisiting Fig. 1, with the idea that mass in range of a weak interaction still somehow might be a useful parameter. A reasonable empirical approach would be to assume some sort of function, and then fit the data with as few free parameters as possible. Here in Fig. 3 is one example:

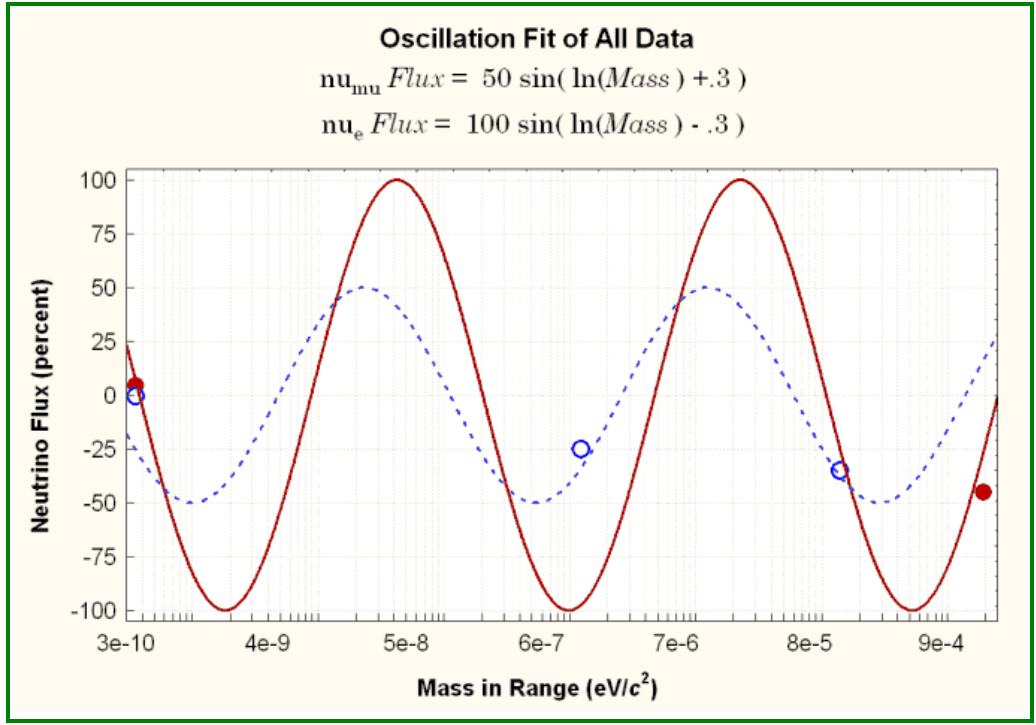

**Figure 3.** An empirical oscillatory fit of $Flux = A\sin(\ln(Mass) + B)$ to the effects in Table 1. There are two free parameters, with different values for each flavor (*B* forced by about 0.05 into algebraic symmetry). Solid points represent $n_e$ data.

If nothing else, the weakness of the usual oscillation formula now becomes manifest: The propagation distance parameter *L* and energy *E* (or momentum *p*) are not free but part of the choice of function to fit, just as were *Flux* and *Mass* in the above. However, the usual oscillation theory cannot explain the effects in Table 1 on the basis of just two neutrino flavors; it requires a hypothetical $\Delta m^2$, as well as one free mixing angle parameter for each of as many as three flavor oscillation modes, for a total of four free parameters, to explain reasonably well the five effects in Table 1.

We note that an oscillation theory matrix with random entries can be used to fit the data as well as any symmetry rationalization [20], reinforcing the idea that it is the number of free parameters, and not the oscillation hypothesis, that makes the oscillation theory satisfactory to those puzzled by neutrinos.

# Another Oscillation Explanation

## *Distance*

Let us look again at the data of Tables 1 and 2, this time taking seriously perhaps the most important of all neutrino properties, their low cross-section of interaction with matter.

Let us assume now that there is no effect of matter, other than a negligible scattering of a few neutrinos, and that the flavors oscillate solely as a function of distance traversed. The fit is shown in Fig. 4:

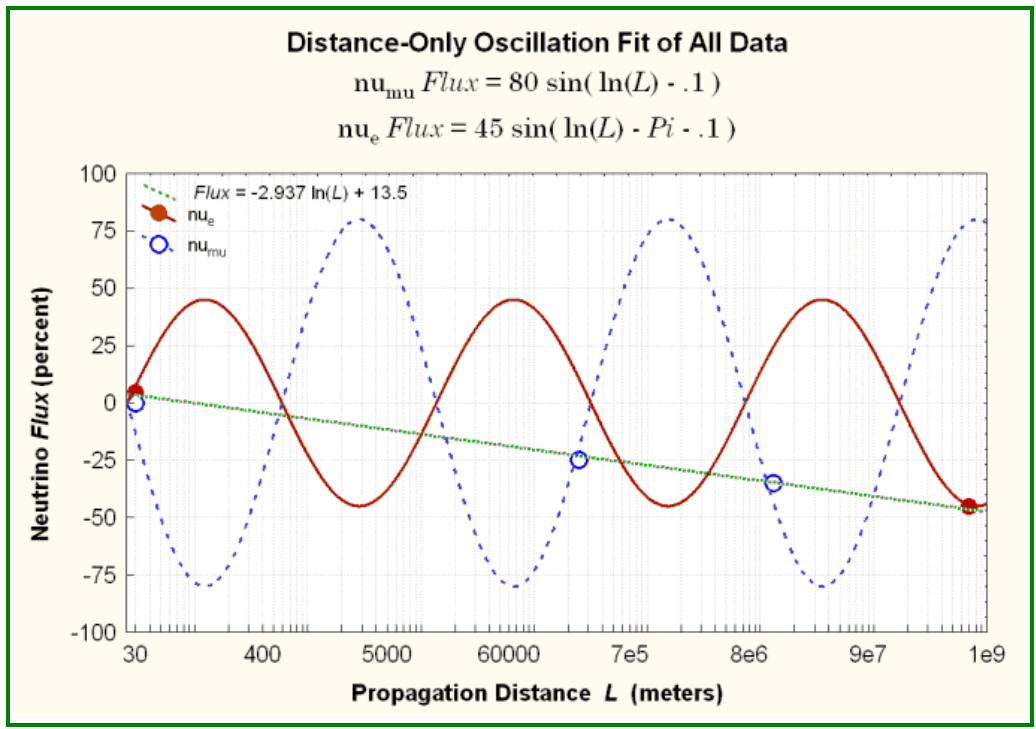

**Figure 4. Empirical oscillatory fit of** $Flux = A\sin(\ln(Distance) + B)$.

Parameters for the separate fits of $n_e$ and $n_m$ in Fig. 4 are not so consistent as those in Fig. 3, where the independent variable was mass in range. This argues for mass as a better independent variable, at least in terms of the empirical functions chosen.

It might seem important in Fig. 4 that the log-linear fit to the data again involves a coefficient of $-3$, as in Eq. (4). However, this is illusory, considering the ranges of the variables plotted: In Fig. 1, we had about 16 natural log units of *Mass* domain and about 4 of *Flux*; in Fig. 4, we had about 17 natural log units of distance $L$. Thus, a linear fit in those domains would be expected to produce about the same value of coefficient no matter where the data fell.

The same criticism may be applied to the usual neutrino oscillation formula, which of course should fit the same data. A far greater density of data points on the mass or distance domain will be needed to show a real confirmation of any theory based on so much freedom to err. This has been recognized, and experiments planned and in progress hopefully will provide a better understanding of neutrinos than now can be asserted.

# A Detector-Centric Explanation

## *Detector Profile*

In an attempt to ascribe the effects of Table 1 entirely to the detector alone, Table 3 was created to represent the cross-section and geometry of the detector. The neutrino capture profile is a cross-section defined as the simple number of countable events per day of operation. The angular subtense of the detector at the neutrino source is used with the capture profile to compute an effective subtense, which may be seen as a measure of the *Flux* capturing efficiency of the detector.

**Table 3. The neutrino data of Table 1 and associated detector profile.**

| Neutrino Flavor | Effect (*Flux*) | Detector | Capture Profile $s$ (day$^{-1}$) | Subtense $q$ at Source (sr) | Effective Subtense $s/q$ (day$^{-1}$sr$^{-1}$) |
|---|---|---|---|---|---|
| electron | −45% | Super-K[a] | 14.7 | $5 \cdot 10^{-9}$ | $2.9 \cdot 10^9$ |
|  |  | SNO[b] | 4.9 | $10^{-9}$ | $4.9 \cdot 10^9$ |
|  | +4.5% | LSND[e] | 1.1 | 0.82 | 1.3 |
| muon | −35% | Super-K[c] | 8.6 | $3.8 \cdot 10^{-5}$ | $2.3 \cdot 10^5$ |
|  | −25% | Super-K[d] | 0.28 | $2 \cdot 10^{-3}$ | 140 |
|  | −25% | LSND[f] | 17.2 | 0.82 | 21 |

[a] Super-K solar: Ref. [2]; 18,464 evt/1258 da; 12.5 m radius @ $10^{11}$ m
[b] SNO solar: Ref. [3]; 1169 evt/240 da; 6 m radius @ $10^{11}$ m.
[c] Super-K atmos: Ref. [13]; 9178FC + 665PC evt/1140 da; 12.5 m radius @ $1.3 \cdot 10^7$ m.
[d] Super-K K2K: Ref. [6]; 28 evt/100 da; 12.5 m radius @ $2.5 \cdot 10^5$ m.
[e] LSND: Ref. [5]; 2000 evt/(6y@300da); 2.8 m radius @ 30 m.
[f] LSND: Ref. [5]; 31000 evt/(6y@300da); 2.8 m radius @ 30 m.

In Table 3, neither the *Capture Profile* nor the *Subtense* alone seemed very useful when inspected graphically; however, the *Effective Subtense* is shown in Figure 5. Once again, a reasonable function, suggested merely by looking at the data, fits reasonably well with only two free parameters.

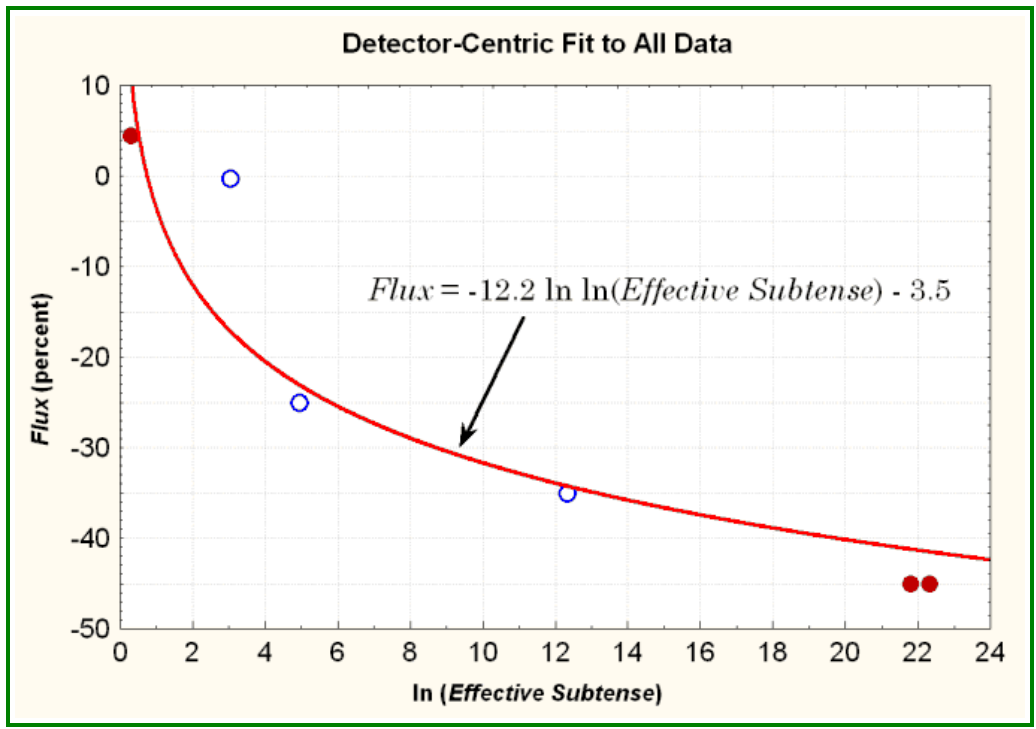

**Figure 5. Empirical fit of Table 1 data by**
$Flux = A \ln\ln(Effective\ Subtense) + B$ **as in Table 3.  Solid points are $n_e$ data.**

## Conclusion

We have no special attachment for any of the empirical fits shown here.  They are, of course, very little related to any theory of neutrino physics.   However, theory building was not the point.   Actually, curve-fitting was not the point, either.   The point was that the data available <u>at</u> <u>present</u> on neutrinos can be represented reasonably well with far fewer parameters than required by the present version of the usual neutrino oscillation theory.

If the deluge of new data expected on neutrinos in the next few years do not fall into close functional correspondence with the data analyzed here, the oscillation theory will have to be extended by addition of yet more free parameters.

The present analysis shows the usual theory to be highly inefficient at representing data already.   We suggest, therefore, that increasing the complexity of the usual theory should be avoided; rather, serious consideration should be given to discarding it in favor of a theory of greater value to physics.

## References

1. J. M. Williams,  "Asymmetric Collision of Concepts:  Why Eigenstates Alone are Not Enough for Neutrino Flavor Oscillations"; poster, *SLAC SSI 28* (2000),  *arXiv*, physics/0007078 (2001).   "Some Problems with Neutrino Flavor Oscillation Theory"; poster in prep, *SLAC SSI 29* (2001), *arXiv*, physics/xxx.

2. S. Fukuda, *et al* (Super-Kamiokande Collaboration), "Solar 8B and *hep* Neutrino Measurements from 1258 Days of Super-Kamiokande Data", *arXiv*, hep-ex/0103032, 2001.

3. Q. R. Ahmad, *et al* (SNO Collaboration), "Measurement of charged current interactions produced by 8B solar neutrinos at the Sudbury Neutrino Observatory", *arXiv*, nucl-ex/0106015, 2001.

4. T. Toshito, "Super-Kamiokande atmospheric neutrino results", *arXiv*, hep-ex/0105023, 2001.

5. A. Aguilar, *et al* (LSND Collaboration), "Evidence for Neutrino Oscillations from the Observation of Nue_bar Appearance in a Numu_bar Beam", *arXiv*, hep-ex/0104049, 2001.

6. S. H. Ahn, et al, "Detection of Accelerator-Produced Neutrinos at a Distance of 250 km", *arXiv*, hep-ex/0103001, 2001.

7. D. R. O. Morrison, "Review of Neutrino Masses and Oscillations", 5th Course Lecture at the 'D. Chalonge' International School of Astrophysics on *Current Topics in Astrofundamental Particles*, Erice, Italy, 7 - 15 September 1996; Proceedings published by Singapore World Science, 1997; CERN online SCAN-9704132 (DM-96-20), 1997.

8. S. M. Bilenky, "Neutrinos", *arXiv*, physics/0103091, 2001.

9. C. W. Kim and A. Pevsner, *Neutrinos in Physics and Astrophysics*, Harwood, 1993.

10. J. Sielaff, "Observation of Nu_tau Charged-Current Interactions", *arXiv*, hep-ex/0105042, 2001.

11. A. Habig, "Discriminating between Nu_mu <--> Nu_tau and Nu_mu <--> Nu_sterile in atmospheric Nu_mu oscillations with the Super-Kamiokande detector", *Proceedings of the ICRC*, 2001 v. **1**, *arXiv*, hep-ex/0106025, 2001.

12. From Super-K online information page at University of Tokyo site, `www-sk.icrr.u-tokyo.ac.jp/doc/sk/super-kamiokande.html` (Downloaded 1999-07-26).

13. S. Fukuda, *et al* (Super-Kamiokande Collaboration), "Tau Neutrinos Favored over Sterile Neutrinos in Atmospheric Muon Neutrino Oscillations", *Physical Review Letters*, v. **85**(19), 3999 - 4003, 2000.

14. F. Vannucci, "Search for Stimulated Neutrino Conversion with an RF Cavity", paper given at Blois, France, *Rencontres de Blois*, June 28 - July 3, *arXiv*, hep-ex/9911025, 1999.

15. V. Barger, *et al*, "Neutrino Decay as an Explanation of Atmospheric Neutrino Observations", *Physical Review Letters*, vol. **82**(13), 2640 - 2643, 1999.

16. F. Vannucci, "Can LSND and Superkamiokande be Explained by Radiative Decays of Nu_mu's?", *arXiv*, hep-ph/9903487, 1999.

17. Y. Grossman and M. P. Worah, "Atmospheric Nu_mu Deficit from Decoherence", SLAC-PUB-7888, *arXiv*, hep-ph/9807511, 1998.

18. M. C. Garcia, *et al*, "Atmospheric Neutrino Observations and Flavor Changing Interactions", *Physical Review Letters*, vol. **82**(16), 3202 - 3205, 1999.

19. R. E. Lopez, *et al*, "Probing Unstable Massive Neutrinos with Current Cosmic Microwave Background Observations", *Physical Review Letters*, vol. **81**(15), 3075 - 3078, 1998.

20. L. Hall, H. Murayama, and N. Weiner, "Neutrino Mass Anarchy", *Physical Review Letters*, vol. **84**(12), 2572 - 2575, 2000.

# Acknowledgements

The manuscript was propared in *MS Word*, and the graphs were created in *Statistica*. Italicized words are used in trade by their respective owners.